# Melting of Spin Ice state through structural disorder in $Dy_2Zr_2O_7$

J. G. A. Ramon[1,*], C. W. Wang[2,#] L. Ishida[1,†], P.L. Bernardo[1], M. M. Leite[3], F. M. Vichi[3], J. S. Gardner[4] and R. S. Freitas[1]

[1] Instituto de Física, Universidade de São Paulo, 05314-970 São Paulo, SP, Brazil.
[2] Neutron Group, National Synchrotron Radiation Research Center, Hsinchu, Taiwan
[3] Instituto de Química, Universidade de São Paulo, São Paulo, SP, Brazil.
[4] Songshan Lake Materials Laboratory, Dongguan, Guangdong 523808, China

Neutron scattering, a.c. magnetic susceptibility and specific heat studies have been carried out on polycrystalline $Dy_2Zr_2O_7$. Unlike the pyrochlore spin ice $Dy_2Ti_2O_7$, $Dy_2Zr_2O_7$ crystallizes into the fluorite structure and the magnetic $Dy^{3+}$ moments randomly reside on the corner-sharing tetrahedral sublattice with non-magnetic Zr ions. Antiferromagnetic spin correlations develop below 10 K but remain dynamic down to 40 mK. These correlations extend over the length of two tetrahedra edges and grow to 6 nearest neighbors with the application of a 20 kOe magnetic field. No Pauling's residual entropy was observed and by 8 K the full entropy expected for a two level system is released. We propose that the disorder melts the spin ice state seen in the chemically ordered $Dy_2Ti_2O_7$ compound, but the spins remain dynamic in a disordered, liquid-like state and do not freeze into a glass-like state that one might intuitively expect.

# I. INTRODUCTION

A magnetic sublattice is frozen in at the high energy scale of material crystallization. The combination of this sublattice, the nature of the spin, and the couplings between spins can thwart the satisfaction of all nearest neighbor interactions simultaneously. This phenomenon is known broadly as geometrically frustrated magnetism [1-3]. Over the past two decades significant advances in our understanding of this magnetism has come from studies on the rare earth series of pyrochlore titanates, $R_2Ti_2O_7$. These model magnets, where the magnetic rare earth ions (R) are distributed on a sublattice of corner-linked tetrahedral have exhibited a remarkable variety of behavior including spin ice (R = Ho, Dy) [4,5], spin liquid (R = Tb) [6,7], partial order (R=Gd) [8,9] and fluctuation-induced magnetic order (R = Er) [10]. Frustration is also relevant in the Spin-1 chain system $CsNiCl_3$ and the formation of stripe-like phases in perovskites [11,12] and plays a critical role outside of magnetism in areas like protein folding and determining the structure of solid nitrogen and water [1,2].

The availability of large single crystals over most of the lanthanide series of pyrochlore titanates, resulted in several transformative works over the past 20 years [7,13-16] However, the open lattice of the oxide pyrochlores, with the chemical formula $R_2M_2O_7$, allows for a large number of combinations of R and M ions. Here the R-site is occupied by a trivalent rare-earth ion with eightfold oxygen co-ordination and the M-site by a tetravalent transition metal ion with sixfold oxygen co-ordination. Both cations are located at the vertices of two distinct networks of corner-sharing tetrahedral and if either R or M is magnetic, frustration can develop. When considering the chemical bonding, the pyrochlore structure can be described as an ordered defect fluorite [17,18], and recent work on several disordered pyrochlores have revealed exotic magnetism [19,20].

Theoretically, it was predicted that bond disorder on the pyrochlore lattice induces spin glass behavior at very small concentrations [21]. Sen and Moessner [22] also predicted frozen spin states in disordered spin glasses. More recently a disorder-induced, quantum-entangled liquid phase was predicted in non-Kramers ions based oxide pyrochlores [23]. Technological applications of pyrochlore materials are extensive ranging from the immobilization of active nuclear waste to high temperature thermal barrier coatings and from luminescence to solid oxide fuel [1,2,17].

A central topic in geometrically frustrated magnetism embraces large, Ising spins on the pyrochlore lattice with ferromagnetic interactions. $Dy_2Ti_2O_7$ is one such pyrochlore oxide, which possesses strong Ising anisotropy along the local <111> directions and ferromagnetic nearest neighbour interactions [5]. Here, the spins on each tetrahedron satisfy the two-in two-out ice rule; this arrangement of magnetic moments can be directly compared to the bonding of hydrogen in hexagonal water ice ($I_h$) and led to the classification of the spin ice ground state almost two decades ago [4,24]. This spin state also possesses Pauling's residual entropy equal to $(R/2)\ln(3/2)$ again similar to $I_h$ [4,25] Recently, experiments have suggested the amount of residual entropy is significantly reduced from the Pauling entropy in $Dy_2Ti_2O_7$ suggesting that the spin system may order if equilibrated properly [26] and consistent with the spin ice model which predicts an ordered ground state at the lowest temperatures [27]. In the seminal works by Snyder *et al.* [28,29], the real and imaginary part of the dynamic susceptibility suggested an exotic crossover from classical to quantum and back to classical relaxation phenomena as spin ice freezes. Further studies [14,28-32] of the dynamical processes in spin ice have confirmed the existence of low temperature spin dynamics and identified some of the processes. Confirming the basic near neighbor and the all-encompassing dipolar spin ice theories [5], these data observe spin flipping through the first excited crystal field level. Reducing the temperature below 2 K, neutron diffraction observes the diffuse scattering that can be described well by a disordered spin system with local ice rules [14]. At these low temperatures, a crossover into the spin ice state occurs with the appearance of a plateau in the [111] isothermal magnetization [33,34] and the reemergence of a thermally activated relaxation process [29, 35] that persist down to 20 mK [28-31,36]. These spin dynamics below 5 K, can be explained in the framework of the creation and propagation of point like defects or emergent magnetic monopoles (a pair of monopoles is produced by the flip of a single spin, giving rise to two neighbor tetrahedra with three-in one-out and three-out one-in configurations) interacting via Coulomb potential [15,32,37,38].

To understand the spin ice state further and in an attempt to influence the magnetic monopoles, researchers have begun to manipulate the chemistry, creating quantum, dilute and stuffed spin ices [16,23,24,39,40]. For example, researchers are looking outside the structure-field map to make new spin ice materials [41], are placing smaller moments in the lattice [42] or looking beyond the typical non-magnetic M-site ions where the structure and availability of crystals is more problematic [20,43,45]. Recently, several groups have been working on geometrically frustrated hafnates and zirconates [20,43-45]. The experimental work is sparse and the low-temperature magnetism for some rare-earth cations is absent. For example; in the non-Kramers pyrochlore $Pr_2Zr_2O_7$, spin-ice-like correlations and strong quantum fluctuations were reported [43,44], but the nature of its ground state is still not fully understood. No magnetic order was observed in the pyrochlore $Nd_2Zr_2O_7$ down to 0.5 K [45] although it has a Curie Weiss temperature of +0.15 K. Similarly, $Tb_2Hf_2O_7$ and $Pr_2Hf_2O_7$ were found not to show long-range magnetic order down to 100 mK [20,46] although $Pr_2Hf_2O_7$ my freeze into a glass state at 90 mK [47]. $Nd_2Hf_2O_7$ was found to exhibit a long-range antiferromagnetic order below $T_N \approx$ 0.55 K with an all-in/all-out spin arrangement [48].

Here, we report low temperature thermodynamic studies and neutron diffraction on dysprosium zirconate, $Dy_2Zr_2O_7$, and discuss its magnetic and structural properties. The trivalent Dy-ion is the only magnetic species in the compound, similar to that in the spin ice, $Dy_2Ti_2O_7$. However, it is known that the small lanthanide elements (including $Dy^{3+}$) favor a disordered fluorite structure, rather than the ordered pyrochlore oxide of spin ice [17,49]. Comparing the two dysprosium compounds, we observe a few thermodynamic similarities, however the spin ice character is completely absent in $Dy_2Zr_2O_7$ and is replaced by spin liquid characteristics. Very dynamic, short-range, antiferromagnetic correlations are observed below 10 K, reminiscent of those found in terbium pyrochlores and described well by the Gardner-Berlinsky (GB) model. [20,50]

## II. EXPERIMENTAL

Polycrystalline powder sample of $Dy_2Zr_2O_7$ was prepared by the sol-gel method. As an alternative to the usual solid state or "shake and bake" reaction, the wet-chemistry, low temperature technique is known to produce excellent atomic level mixing, greater control over particle morphology and size [51]. Dysprosium oxide, $Dy_2O_3$, and tetrabutyl zirconate, $C_{16}H_{36}O_4Zr$, were employed as precursors of the dissolutions of the cations $Dy^{3+}$ and $Zr^{4+}$ respectively. The solution was stirred until it changes into a gel, and finally the resulting gel is calcinated at 1100° C for 24 hours, as detailed in [51,52]. The same method was employed for preparing $Dy_2Ti_2O_7$, and the non-magnetic $Lu_2Zr_2O_7$ used for specific heat analysis. X-ray powder diffraction experiments were performed using a Shimadzu XRD-7000 diffractometer. The Bragg-Brentano geometry and Cu K$\alpha_1$ radiation (1.5406 Å) was used. The crystal structure was characterized by performing a least-squares Rietveld refinement of the powder neutron diffraction data, using the FULLPROF software suite [53] and the graphical interface WinPLOTR [54].

Magnetic measurements were carried on using the superconducting quantum interference device (SQUID, Quantum Design) down to 1.8 K. The ac magnetic susceptibility ($\chi_{ac}$) data were collected on an adiabatic demagnetization refrigerator (ADR, Cambridge Cryogenics) with an amplitude and phase compensator circuit in the frequency range from 10 Hz to 10 kHz in a modulation field of 0.5 Oe. Specific heat experiments were conducted using the calorimeter insert of a Physical Property Measurement System (PPMS, Quantum Design) operating with a dilution refrigerator to reach 50 mK. To reduce the effect of the high neutron absorption cross section of $^{161}$Dy ($600*10^{-28}$ m$^2$) and $^{163}$Dy ($2840*10^{-28}$ m$^2$), $^{162}$Dy$_2$Zr$_2$O$_7$ was also synthesized by the sol-gel method using the 98% enriched $^{162}$Dy$_2$O$_3$ for neutron studies. Data were collected at the high-intensity neutron diffractometer, WOMBAT [55] at the Australian Nuclear Science and Technology Organization (ANSTO, Sydney) using 4.744 Å neutrons. The 300 mg sample was placed in an oxygen-free copper can and mounted to the end of a dilution cryostat to reach a temperature of 40 mK.

## III. RESULTS AND DISCUSSION

X-ray powder diffraction found all samples to be single phase without any detectable impurity. Modeling the data with the cubic space group, $Fm\bar{3}m$, consistent with the defect-fluorite structure resulted in an excellent fit as shown in the bottom inset of figure 1. The data was not described as well with the oxide pyrochlore structure and space group, $Fd\bar{3}m$. Neutron diffraction on $^{162}$Dy$_2$Zr$_2$O$_7$ confirmed the X-ray Rietveld refinement and found a cubic lattice constant of value $a$ = 5.238 (2) Å. This structure has one cation site and one oxygen site that is 7/8 occupied. In the defect fluorite structure, 8 O atoms locally form a perfect cube around both metal ions. This is significantly different from the local coordination of the rare earth site in the oxide pyrochlore structure, where 6 oxygen atoms form a puckered hexagon with the other two oxygen atoms forming one of the shortest rare-earth oxygen bonds known, along the local <111> direction. This change in the local environment should result in a strong modification to the CEF states.

Fitting the Curie-Weiss law to the inverse susceptibility of Dy$_2$Zr$_2$O$_7$ in the paramagnetic regime above 100 K, yields an antiferromagnetic Curie-Weiss temperature $\Theta_{CW}$ = -10.8 K. However, if we fit in the linear regime between 10 K and 100 K, as shown in figure 1, this changes to $\Theta_{CW}$ = -3.2 K. A ferromagnetic $\Theta_{CW}$ = +0.5 K was reported for the spin ice Dy$_2$Ti$_2$O$_7$ [5]. The top inset of figure 1 shows the saturation magnetization as a function of the applied field of Dy$_2$Zr$_2$O$_7$. The saturation moment at 7 T is close to 5 $\mu_B$/Dy ion similar to Dy$_2$Ti$_2$O$_7$, which is the half the free ion value and due to its crystal field anisotropy [56]. These data indicate that Dy$_2$Zr$_2$O$_7$ has dominant antiferromagnetic interactions and easy axis anisotropy.

Measurement of the real part of the ac magnetic susceptibility, $\chi'_{ac}$, reveal a frequency dependent maximum at T' ≈ 1 K, as shown in figure 2. The shape and height of these maxima are similar to those observed in the pyrochlore spin ice Dy$_2$Ti$_2$O$_7$ [28,30] but at a slightly lower temperature. The drop in $\chi'_{ac}$ after the maximum indicates that the spins response is slow and they are not able to keep up with the time-varying magnetic field. Contrary to what is observed in $Dy_2Ti_2O_7$ the values of $\chi'_{ac}$ does not vanish below 0.5 K, revealing an incomplete spin freezing of the system and the ubiquitous presence of persistent spin dynamics seen in many geometrically frustrated magnets. The maximum close to 1 K shifts toward higher temperatures and becomes broader as the frequency of the ac measurement increases. We can characterize the dynamics of $Dy_2Zr_2O_7$ by fitting the frequency, $f$, versus the temperature, $T'$, of the maximum in $\chi'_{ac}$ to an Arrhenius law $f = f_0\ exp(-E_b/k_BT')$, where $E_b$ is the energy barrier. The entire temperature and frequency range measured (see inset of figure 2) is described well by the Arrhenius law revealing a single characteristic relaxation time of $\tau_0 = 1/2\pi f_0 = 4.5 \times 10^{-5}$ s and an energy barrier $E_b = 8$ K. These values are close to parameters obtained in $Dy_2Ti_2O_7$ below 1 K, which are an energy barrier $E_b \approx 10$ K and a characteristic relaxation time on the scale of $10^{-7}$ s [28,30,38,57].

Specific heat experiments were performed down to 70 mK for both $Dy_2Zr_2O_7$ and $Dy_2Ti_2O_7$. Figure 3 shows the temperature dependence of the total specific heat (C), the lattice term $C_p$ accounted by measuring the non-magnetic fluorite $Lu_2Zr_2O_7$, and the non-negligible nuclear contribution [58,59] at the lowest temperatures. This nuclear term arises from the nuclear magnetic moment, I = 5/2 of two isotopes, $^{161}$Dy and $^{163}$Dy, with quadrupole and hyperfine interactions as discussed by Henelius *et al.* [58]. The specific heat data of $Dy_2Zr_2O_7$ shows a broad peak at about 2 K shares some resemblance to the classical spin-ice $Dy_2Ti_2O_7$ [4], and is associated with static short-range correlations. After isolating the electronic specific heat $C_e$, The recovered entropy $\Delta S_e(T)$ was obtained by integrating $C_e(T)/T$ from the minimum measured to the temperature T. Figure 4 shows that the values of $\Delta S_e(T)$ for the $Dy_2Zr_2O_7$ is close to the expected Rln(2); for a system with only two discrete orientations, at approximately 8 K. This indicates that the residual entropy left as T goes to zero is small, if not negligible, significantly less than the (R/2)ln(3/2) found in spin ice ($Dy_2Ti_2O_7$) and water [60] and understood by Pauling for water ice [25].

A well characterized isotopically enriched, powder sample was used for neutron diffraction investigations. The bulk thermodynamic properties and the X-ray powder diffraction pattern of $^{162}Dy_2Zr_2O_7$ were consistent with our other samples. Neutron diffraction confirms that there is no long-range order down to 39 mK, but revealed broad diffuse scattering below 10 K, associated with the correlated spins. Data measured at 10 K revealed a relatively flat background and sharp, resolution limited, Bragg peaks associated with the crystalline structure (not shown). These data were used as a background dataset and subtracted from the lower temperature data to enhance the magnetic scattering.

Magnetic diffraction data sets at 40 mK are shown in figure 5. The net intensity has been corrected for the |**Q**|-dependence due to the $Dy^{3+}$ magnetic form factor so that models of possible spin structures can be compared. Although the |**Q**| range and the statistical significance of the data is not of the quality necessary to model the diffuse scattering, several observations can immediately be drawn from these data. First, the broad, liquid like, distribution of magnetic scattering centered at roughly 1.2 Å$^{-1}$ is characteristic of antiferromagnetically coupled Ising spins on the corner sharing tetrahedral lattice [6,7,9,20,50,61-67]. Second, the lack of forward scattering at low |**Q**| indicating the absence of ferromagnetic correlations and consistent with our negative Curie-Weiss temperature. Finally, the data appears to reach a second maximum at ~ 2.3 Å$^{-1}$. All these observations are reminiscent of those from the pyrochlore $Tb_2Mo_2O_7$ [63] where an antiparallel alignment of adjacent spins is observed on both the R and M sublattice and the disordered pyrochlore $CsNiCrF_6$ [67]. The width of the peak at the

antiferromagnetic correlations wavevector of 1.2 Å$^{-1}$ suggests the mean correlation length is that of the two M-M bonds on the fluorite structure (5.24 Å). To understand the correlations in Dy$_2$Zr$_2$O$_7$ further, we also plot the calculated powder average structure factor for spin ice [5,24] and that for near-neighbor antiferromagnetic correlations on the pyrochlore lattice [50]. The metal-ion tetrahedra still exist, albeit disordered, in this fluorite lattice and with nearest neighbor correlations only it is not unreasonable to begin with these models. The near-neighbor antiferromagnetic correlations on the pyrochlore lattice, so-called Gardner-Berlinsky (GB) model was recently used to describe the magnetic correlations in the disordered pyrochlore Tb$_2$Hf$_2$O$_7$ [20] and originally used to model the spin liquid, Tb$_2$Ti$_2$O$_7$ [50]. Clearly spin ice correlations are not present in this Dy compound. At first glance, the data is described well by the GB model, however the poor statistics makes it difficult to precisely model the high $|\mathbf{Q}|$ data, where we see a peak at 2.3 Å$^{-1}$ and the model predicts the second peak closer to 3.0 Å$^{-1}$. Considering the extensive disorder in the fluorite structure, the powder averaged data and the limited amount of momentum space covered no further attempts were made to model these data and adding to the number of variables.

To stress the surprising similarity in the low temperature relaxation processes within the ferromagnetic, pyrochlore, spin ice and this antiferromagnetic, Dy$^{3+}$-based fluorite, the temperature dependence of the characteristic relaxation time is plotted in figure 6. Taken from the dynamic susceptibility measured below 2 K, where the spatial spin-spin correlations are known to exist from our neutron diffraction studies (figure 5), the strong temperature dependence and the characteristic time scale matches that of Dy$_2$Ti$_2$O$_7$ [29,57]. This similarity reached deep into the spin ice phase of Dy$_2$Ti$_2$O$_7$ where monopole excitations are created and move throughout the lattice. This behavior of the time evolution of our system is also analogue to the exhibited in the antiferromagnetic Er$^{3+}$-based spinels CdEr$_2$X$_4$ (X=Se, S) which present monopole populations [68,69]. At higher temperatures, no quantum tunneling regime was seen in Dy$_2$Zr$_2$O$_7$, as reported for Dy$_2$Ti$_2$O$_7$ [36,37].

In presence of magnetic field the shape of the low-angle scattering visibly changed. When 10 kOe are applied, as shown in figure 7, the broad scattering sharpens up, but remains centered at 1.15(7) Å$^{-1}$. Fitting this and other data results in an order parameter like curve shown in the inset to figure 7. Here we plot the correlation length from the full-width, half-maximum of the broad diffuse scattering. The short-range, spin-spin correlations lengthen in a field, but appear to saturate above 15 kOe. This may be a plateau and more studies are necessary. Above 15 kOe the correlations extend to 24 Å, approximately 6 nearest neighbor lengths or just short of 5 unit cells.

## IV. CONCLUSION

The oxide pyrochlore structure can be considered an ordered $2 \times 2 \times 2$ superstructure of the disordered defect fluorite. We have used wet chemistry and low temperature annealing to produce a well-mixed, disordered fluorite, Dy$_2$Zr$_2$O$_7$. Here we expect the metal ions to decorate two interpenetrating sublattices of corner-sharing tetrahedral, with a statistical distribution of Dy-Dy, Zr-Zr and Dy-Zr bonds all with the same bond length, but only Dy$^{3+}$ is magnetic. Trivalent dysprosium pyrochlore oxides (Dy$_2$Ti$_2$O$_7$ [4,5] and Dy$_2$Sn$_2$O$_7$ [70]) have both shown spin ice character. Here we have shown that Dy$_2$Zr$_2$O$_7$, with its disordered lattice, but similar moment size, near neighbor distance, and spin anisotropy is a novel spin liquid candidate. Neutron diffraction and specific heat revealed short range spin-spin correlations below 2 K or $|\Theta_{CW}|$, and no long-ranged ordered down to 40 mK.

These spin correlations are antiferromagnetic, very short range, extending over next-nearest neighbors only. Unlike the classical spin ices, susceptibility and specific heat indicate the presence of significant spin dynamics at the lowest temperatures and absence of residual entropy. Below 2 K, neutron diffraction reveals significant spin-spin spatial correlations, or entanglement, whilst the time correlations measured by $\chi'_{ac}$, indicates that the system is very dynamic. These two ingredients are essential for the identification of a quantum spin liquid and we propose that this fluorite indeed possesses such a ground state.

These studies have shown that a significant amount of disorder can lead to a dynamic ground state when combined with frustration at low temperatures. Other geometrically frustrated magnets also show that disorder plays a similar role. Here we show that disorder melts the Kramers spin ice. With the availability of zirconate single crystals [71] and the possibility of solid solution between the spin ice and spin liquid end members this family seems to be an excellent platform for further investigations, probing the role of disorder on the spin dynamics of pyrochlores, the melting of spin ice and the propagation of monopoles as disorder is induced to the coulomb phase.

## ACKNOWLEDGMENTS

The authors (JSG and CCW) would like to thank the Ministry of Science and Technology (MOST) of the Taiwan for financially supporting this research under Contracts No. MOST 120−2739−M−213−001−MY3 and the support of SIKA-Neutron Program of MOST, and ANSTO staff for neutron beamtime assistance. RSF acknowledges FAPESP Grant No. 2015/16191-5 and CNPq Grant No. 429511/2018-3. FMV acknowledges FAPESP Grant 2011/19941-4.

† Present address: Instituto de Física "Gleb Wataghin", UNICAMP, Campinas, São Paulo 13083-970, Brazil.

[*]gtvar@if.usp.br
[#] wang.cw@nsrrc.org.tw

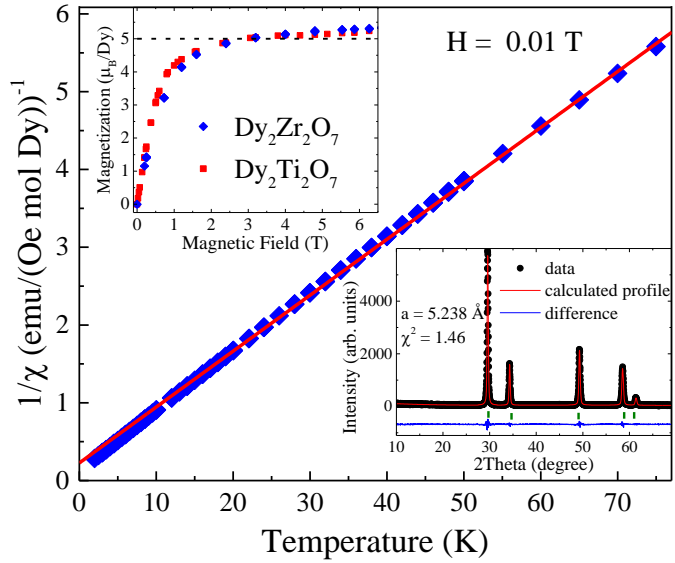

Figure 1. Temperature dependence of the inverse susceptibility and Curie-Weiss fit to the lowest temperature. Top inset: saturation magnetization as a function of the applied field shows a saturation moment of approximately 5 $\mu_B$/Dy ion (dashed line). Bottom inset: x-ray powder diffraction and calculated profile data for $Dy_2Zr_2O_7$ at room temperature. Peak positions of the fluorite structure are marked with small vertical lines.

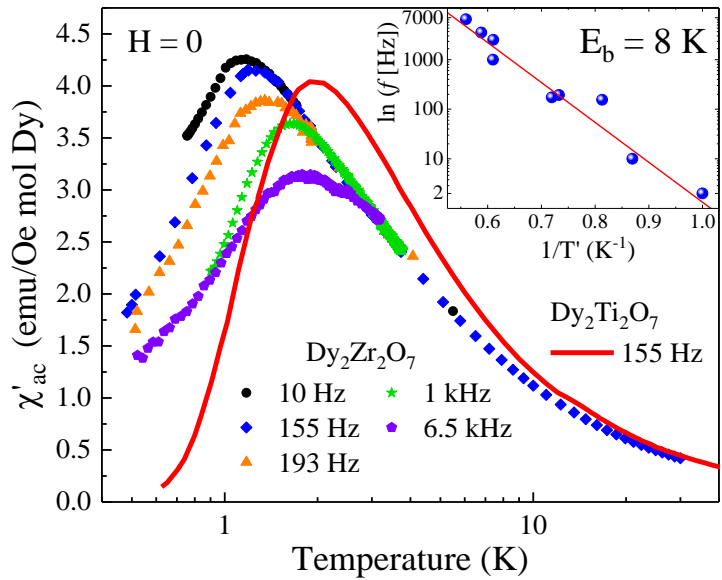

Figure 2. Real part of the ac magnetic susceptibility $\chi'_{ac}$ versus temperature for different frequencies in zero magnetic field for $Dy_2Zr_2O_7$ and $Dy_2Ti_2O_7$ (red solid line). Inset shows the frequency versus the inverse of the temperature of the maximum in $\chi'_{ac}$ with a fit to the Arrhenius law.

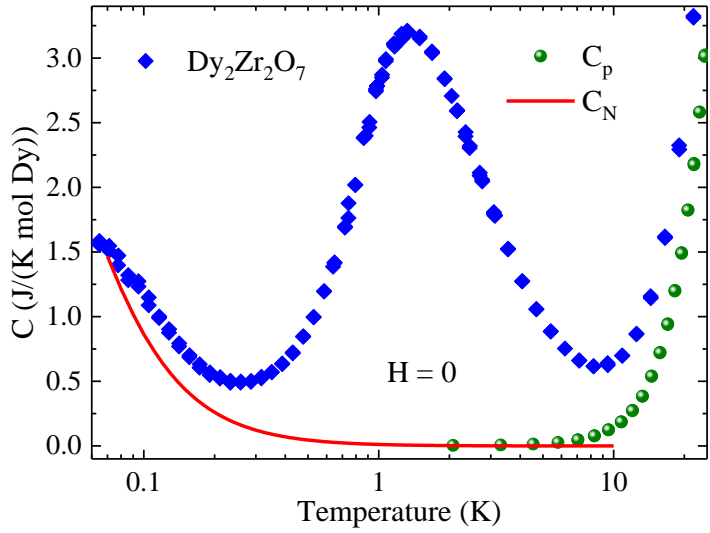

Figure 3. Temperature dependence of the total specific heat of $Dy_2Zr_2O_7$ at zero field. The lattice term ($C_P$) was obtained from measurements on non-magnetic $Lu_2Zr_2O_7$, and the nuclear specific heat contribution ($C_N$) computed as explained in the text.

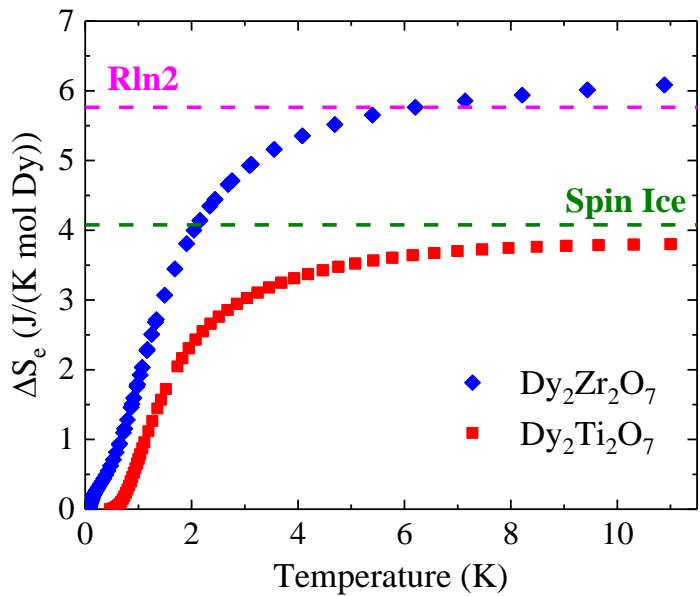

Figure 4. The recovered electronic entropy $\Delta S_e$ as a function of the temperature for $Dy_2Zr_2O_7$ and $Dy_2Ti_2O_7$. The dashed lines denote the expected values for Ising spins (Rln2) and spin ices.

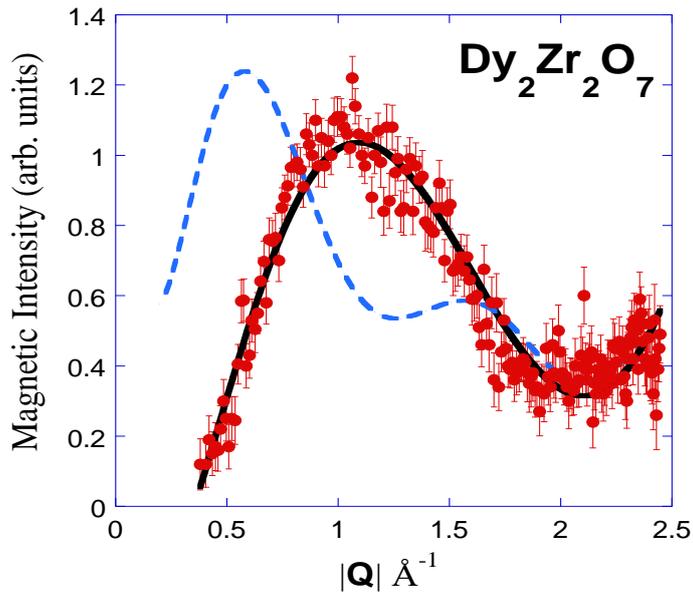

Figure 5. Magnetic neutron scattering for $Dy_2Zr_2O_7$ at 40mK, after a data set at 10 K was subtracted to remove the non-magnetic background, including that from the crystalline structure. Data (red circles) are plotted against the powder averaged dipolar spin ice model (dashed line) and the Gardner-Berlinsky model (solid line).

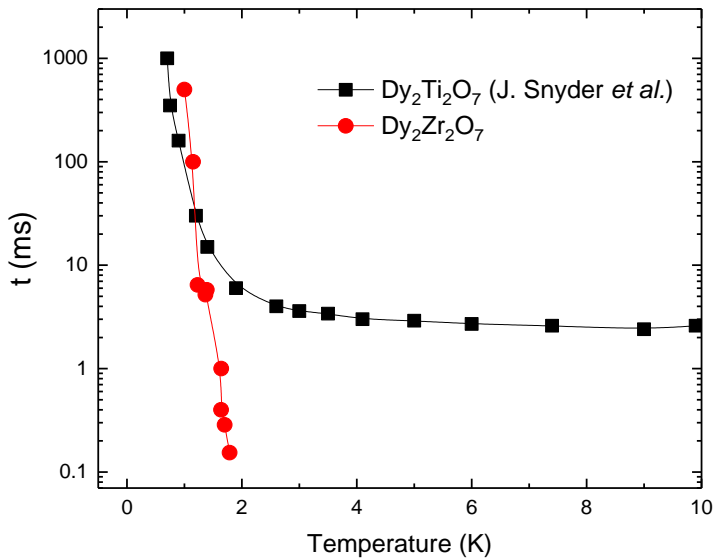

Figure 6. The characteristic spin relaxation time as a function of temperature in $Dy_2Zr_2O_7$ and $Dy_2Ti_2O_7$ (taken from Snyder et al. [66]). The spin relaxation time within $Dy_2Zr_2O_7$, closed circles, increases rapidly below 2K, and match those in the spin ice phase of $Dy_2Ti_2O_7$.

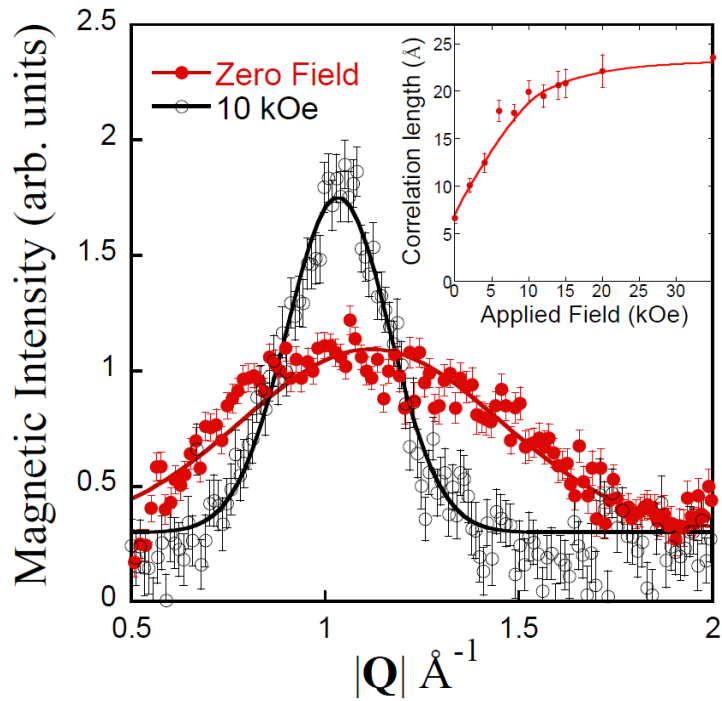

Figure 7. Spatial spin-spin correlations as a function of applied magnetic field. Main panel: The magnetic diffraction at 0 and 10 kOe, Gaussian fits with a common instrumental background are used to describe the data. Inset: the field dependence of the correlation length determined from the Gaussian width of the broad maxima seen around $|Q| \approx 1.15$ Å$^{-1}$.

1. H. T. Diep, *Frustrated Spin Systems* (World Scientific, Singapore, 2013).

2. J. S. Gardner, M. J. P. Gingras, and J. E. Greedan, Rev. Mod. Phys. **82**, 53 (2010).

3. J. S. Gardner, J. Phys: Cond. Matter **23,** 160301 (2011).

4. A. P. Ramirez, A. Hayashi, R. J. Cava, R. Siddharthan, and B. S. Shastry, Nature (London) **399**, 333 (1999).

5. B. C. den Hertog and M. J. P. Gingras, Phys. Rev. Lett. **84**, 3430 (2000).

6. J. S. Gardner S. R. Dunsiger, B. D. Gaulin, M. J. P. Gingras, J. E. Greedan, R. F. Kiefl, M. D. Lumsden, W. A. MacFarlane, N. P. Raju, J. E. Sonier *et al*., Phys. Rev. Lett. **82**, 1012 (1999).

7. J. S. Gardner, A. Keren, G. Ehlers, C. Stock, Eva Segal, J. M. Roper, B. Fåk, M. B. Stone, P. R. Hammar, D. H. Reich *et al*., Phys. Rev. B **68**, 180401 (2003).

8. J. D. M. Champion, A. S. Wills, T. Fennell, S. T. Bramwell, J. S. Gardner, and M. A. Green *et al*., Phys. Rev. B **64,** 140407 (2001).

9. J. R. Stewart, G. Ehlers, A. S. Wills, S. T. Bramwell, and J. S. Gardner, J. Phys: Cond. Matter **16**, L321 (2004).

10. L. Savary, K. A. Ross, B. D. Gaulin, J. P. C. Ruff, and L. Balents, Phys. Rev. Lett. **109**, 167201 (2012).

11. M. Kenzelmann, R. A. Cowley, W. J. L. Buyers, R. Coldea, J. S. Gardner, M. Enderle, D. F. McMorrow, and S. M. Bennington, Phys. Rev. Lett. **87**, 017201 (2001).

12. V. J. Emery, S. A. Kivelson, and J. M. Tranquada, Proc. Natl. Acad. Sci. U S A **96**, 8814 (1999).

13. J. S. Gardner, B. D. Gaulin, and D. McK. Paul, J. Cryst. Growth **191**, 740 (1998).

14. T. Fennell, O. A. Petrenko, B. Fåk, S. T. Bramwell, M. Enjalran, T. Yavors'kii, M. J. P. Gingras, R. G. Melko, and G. Balakrishnan, Phys. Rev. B **70**, 134408 (2004).

15. C. Castelnovo, R. Moessner, and S. L. Sondhi, Nature Letters **451**, 42 (2008).

16. S. R. Giblin, S. T. Bramwell, P. C.W. Holdsworth, D. Prabhakaran, and I. Terry, Nat. Phys. **7**, 252 (2011).

17. M. A. Subramanian, G. Aravamudan, and G. V. Subba Rao, Prog. Solid State Chem. **15**, 55 (1983).

18. C. Karthik, T. J. Anderson, D. Gout, and R. Ubic, J. Solid State Chem. **194**, 168 (2012).

19. G. C. Lau, R. S. Freitas, B. G. Ueland, B. D. Muegge, E. L. Duncan, P. Schiffer, and R. J. Cava, Nat. Phys. **2**, 249 (2006).

20. R. Sibille, E. Lhotel, M. C. Hatnean, G. J. Nilsen, G. Ehlers, A. Cervellino, E. Ressouche, M. Frontzek, O. Zaharko, V. Pomjakushin *et al*., Nat. Commun. **8**, 892 (2017).

21. T. E. Saunders and J. T. Chalker, Phys. Rev. Lett. **98**, 157201 (2007).

22. A. Sen and R. Moessner, Phys. Rev. Lett. **114**, 247207 (2015).

23. L. Savary and L. Balents, Phys. Rev. Lett. **118**, 087203 (2017).

24. S. T. Bramwell and M. J. P. Gingras, Science **294**, 1495 (2001).

25. L. Pauling, J. Am. Chem. Soc. **57**, 2680 (1935).

26. D. Pomaranski, L. R. Yaraskavitch, S. Meng, K. A. Ross, H. M. L. Noad, H. A. Dabkowska, B. D. Gaulin, and J. B. Kycia, Nat. Phys. **9**, 353 (2013).

27. R. G. Melko, B. C. den Hertog, and M. J. P. Gingras, Phys. Rev. Lett. **87** 067203 (2001).

28. J. Snyder, J. S. Slusky, R. J. Cava, and P. Schiffer, Nature **413**, 48-51 (2001).

29. J. Snyder, B. G. Ueland, J. S. Slusky, H. Karunadasa, R. J. Cava, and P. Schiffer, Phys. Rev. B **69**, 064414 (2004).

30. K. Matsuhira, Y Hinatsu, and T Sakakibara, J. Phys: Cond. Matter **13**, L737 (2001).

31. J. Lago, S. J. Blundell and C. Baines, J. Phys: Cond. Matter **19**, 326210 (2007).

32. D. J. P. Morris, D. A. Tennant, S. A. Grigera, B. Klemke, C. Castelnovo, R. Moessner, C. Czternasty, M. Meissner, K. C. Rule, J.-U. Hoffmann *et al*. Science **326**, 411 (2009).

33. A. L. Cornelius and J. S. Gardner, Phys. Rev. B **64,** 060406**(R)** (2001).

34. O. A. Petrenko, M. R. Lees, and G. Balakrishnan, J. Phys: Cond. Matter **23**, 164218 (2011).

35. J. P. Clancy, J. P. C. Ruff, S. R. Dunsiger, Y. Zhao, H. A. Dabkowska, J. S. Gardner, Y. Qiu, J. R. D. Copley, T. Jenkins, and B. D. Gaulin. Phys. Rev. B **79,** 014408 (2009).

36. G. Ehlers, A. L. Cornelius, M. Orendác, M. Kajnaková, T. Fennell, S. T. Bramwell, and J. S. Gardner, J. Phys: Cond. Matter **15**, L9 (2003).

37. L. D. C. Jaubert and P. C. W. Holdsworth, Nat. Phys. **5**, 258 (2009).

38. L. D. C. Jaubert and P. C. W. Holdsworth, J. Phys: Cond. Matter **23**, 164222 (2011).

39. C. R. Wiebe, J. S. Gardner, S. -J. Kim, G. M. Luke, A. S. Wills, B. D. Gaulin, J. E. Greedan, I. Swainson, Y. Qiu, and C.Y. Jones, Phys. Rev. Lett. **93**, 076403 (2004).

40. J. S. Gardner, A. L. Cornelius, L. J. Chang, M. Prager, Th. Brückel, and G. Ehlers, J. Phys: Cond. Matter **17**, 7089 (2005).

41. H. D. Zhou, S. T. Bramwell, J. G. Cheng, C. R. Wiebe, G. Li, L. Balicas, J. A. Bloxsom, H. J. Silverstein, J. S. Zhou, J. B. Goodenough *et al.*, Nat. Commun. **2**, 478 (2011).

42. P. M. Sarte, A. A. Aczel , G. Ehlers, C. Stock, B. D. Gaulin, C. Mauws, M. B. Stone, S. Calder, S. E. Nagler, J. W. Hollett *et al.*, J. Phys: Cond. Matter **29**, 45LT01 (2017).

43. K. Kimura, S. Nakatsuji, J. -J. Wen, C. Broholm, M. B. Stone, E. Nishibori, and H. Sawa, Nat. Commun. **4**, 1934 (2013).

44. J. -J. Wen, S. M. Koohpayeh, K. A. Ross, B. A. Trump, T. M. McQueen, K. Kimura, S. Nakatsuji, Y. Qiu, D. M. Pajerowski, J. R. D. Copley *et al.*, Phys. Rev. Lett. **118** 107206 (2017).

45. M. C. Hatnean, M. R. Lees, O. A. Petrenko, D. S. Keeble, G. Balakrishnan, M. J. Gutmann, V. V. Klekovkina, and B. Z. Malkin, Phys. Rev. B **91,** 174416 (2015).

46. V. K. Anand, A. T. M. N. Islam, A. Samartzis, J. Xu, N. Casati, and B. Lake, J. Cryst. Growth **498**, 124 (2018).

47. V. K. Anand, L. Opherden, J. Xu, D. T. Adroja, A. T. M. N. Islam, T. Herrmannsdörfer, J. Hornung, R. Schönemann, M. Uhlarz, H. C. Walker *et al.*, Phys. Rev. B **94,** 144415 (2016).

48. V. K. Anand, A. K. Bera, J. Xu, T. Herrmannsdörfer, C. Ritter, and B. Lake, Phys. Rev. B **92**, 184418 (2015).

49. M. J. D. Rushton, R. W. Grimes, C. R. Stanek, and S. Owens, J. of Mater. Res. **19,** 1603 (2004).

50. J. S. Gardner, B. D. Gaulin, A. J. Berlinsky, P. Waldron, S. R. Dunsiger, N. P. Raju, and J. E. Greedan, Phys. Rev. B **64**, 224416 (2001).

51. A. Garbout, S. Bouattour, and A.W. Kolsi. J. Alloy Compd. **469**, 229 (2009).

52. J. G. A. Ramon, MSc. Dissertation (Universidade de São Paulo, Brazil, 2015).

53. J. Rodríguez-Carvajal, Physica B **192**, 55 (1993).

54. T. Roisnel and J. Rodríguez-Carvajal, Mater. Sci. Forum **378-381**, 118 (2001).

55. A. J. Studer, M. E. Hagen1, T. J. Noakes, Physica B **385-386**, 1013 (2006).

56. H. Fukazawa, R. G. Melko, R. Higashinaka, Y. Maeno, and M. J. P. Gingras, Phys. Rev. B **65**, 054410 (2002).

57. L. R. Yaraskavitch, H. M. Revell, S. Meng, K. A. Ross, H. M. L. Noad, H. A. Dabkowska, B. D. Gaulin, and J. B. Kycia, Phys. Rev. B **85,** 020410(R) (2012).

58. P. Henelius, T. Lin, M. Enjalran, Z. Hao, J. G. Rau, J. Altosaar, F. Flicker, T. Yavors'kii, and M. J. P. Gingras, Phys. Rev. B **93,** 024402 (2016).

59. S. R. Giblin, M. Twengström, L. Bovo, M. Ruminy, M. Bartkowiak, P. Manuel, J. C. Andresen, D. Prabhakaran, G. Balakrishnan, E. Pomjakushina *et al*., Phys. Rev. Lett. **121**, 067202 (2018).

60. W. F. Giauque and M. F. Ashley, Phys. Rev. **43,** 81 (1993).

61. S. Guitteny, S. Petit, E. Lhotel, J. Robert, P. Bonville, A. Forget, and I. Mirebeau, Phys. Rev. B **88,** 134408 (2013).

62. I. Mirebeau, A. Apetrei, J. Rodríguez-Carvajal, P. Bonville, A. Forget, D. Colson, V. Glazkov, J. P. Sanchez, O. Isnard, and E. Suard, Phys. Rev. Lett. **94**, 246402 (2005).

63. G. Ehlers, J. E. Greedan, J. R. Stewart, K. C. Rule, P. Fouquet, A. L. Cornelius, C. Adriano, P. G. Pagliuso, Y. Qiu, and J. S. Gardner, Phys. Rev. B **81**, 224405 (2010).

64. J. E. Greedan, J. N. Reimers, C. V. Stager, and S. L. Penny, Phys. Rev. B **43,** 5682 (1991).

65. A. Apetrei, I. Mirebeau, I. Goncharenko, D. Andreica, and P. Bonville, J. Phys: Cond. Matter **19,** 145214 (2007).

66. H. D. Zhou, C. R. Wiebe, A. Harter, N. S. Dalal, and J. S. Gardner, J. Phys: Cond. Matter **20,** 325201 (2008).

67. M. J. Harris, M. P. Zinkin, Z. Tun, B. M. Wanklyn, and I. P. Swainson, Phys. Rev. Lett. **73**, 189 (1994).

68. Shang Gao, O. Zaharko, V. Tsurkan, L. Prodan, E. Riordan, J. Lago, B. Fåk, A. R. Wildes, M. M. Koza, C. Ritter *et al*., Phys. Rev. Lett. **120**, 137201 (2018).

69 G. C. Lau, R. S. Freitas, B. G. Ueland, P. Schiffer, and R. J. Cava, Phys. Rev. B **72,** 054411 (2005).

70. K. Matsuhira, Y. Hinatsu, K. Tenya, H. Amitsuka, and T. Sakakibara, J. Phys. Soc. Jpn. **71**,1576 (2002).

71. M. C. Hatnean, M.R. Lees, and G. Balakrishnan, J. Cryst. Growth **418**, 1 (2015).